# Enhancing lithological interpretation from petrophysical well log of IODP expedition 390/393 using machine learning

*Raj Sahu*[1], Saumen Maiti[1], IIT (ISM) Dhanbad*
[1]Machine Learning and Geocomputing Laboratory, Department of Applied Geophysics, IIT (ISM) Dhanbad

**Summary**

Enhanced lithological interpretation from well logs plays a key role in geological resource exploration and mapping, as well as in geo-environmental modeling studies. Core and cutting information is useful for making sound interpretations of well logs; however, these are rarely collected at each depth due to high costs. Moreover, well log interpretation using traditional methods is constrained by poor borehole conditions. Traditional statistical methods are mostly linear, often failing to discriminate between lithology and rock facies, particularly when dealing with overlapping well log signals characterized by the structural and compositional variation of rock types. In this study, we develop multiple supervised and unsupervised machine learning algorithms to jointly analyze multivariate well log data from Integrated Ocean Drilling Program (IODP) expeditions 390 and 393 for enhanced lithological interpretations. Among the algorithms, Logistic Regression, Decision Trees, Gradient Boosting, Support Vector Machines (SVM), k-Nearest Neighbors (KNN), and Multi-Layer Perceptron (MLP) neural network models, the Decision Tree and Gradient Boosting models outperformed the others, achieving an accuracy of 0.9950 and an F1-score of 0.9951. While unsupervised machine learning (ML) provides the foundation for cluster information that inherently supports the classification algorithm, supervised ML is applied to devise a data-driven lithology clustering mechanism for IODP datasets. The joint ML-based method developed here has the potential to be further explored for analyzing other well log datasets from the world's oceans.

**Introduction**

Machine learning has emerged as a powerful tool for subsurface characterization, offering data-driven approaches to enhance geological interpretation (Qi et al., 2021). In marine drilling programs, such as the International Ocean Discovery Program (IODP), vast amounts of core and well-log data are collected. However, traditional interpretation methods often face challenges due to data incompleteness, lithological complexity, and subjective biases (Larsen et al., 2018). The integration of machine learning with these datasets can potentially improve lithological classification and facies prediction, leading to a more accurate understanding of the subsurface (Zhao et al., 2020).

The present study focuses on well log data analysis from IODP Expeditions 390/393, which investigated the South Atlantic Ocean's crustal and sedimentary processes (Hodell et al., 2022). Figure 1 illustrates the bathymetric map of the expedition sites, including U1556, U1557, and U1561, where seismic reflection profiles were acquired. We have considered well U1556B for our present analysis.

IODP Expeditions 390/393 provide a valuable dataset for investigating crustal and sedimentary processes. However, data limitations, such as missing intervals and sparse lithological labels, make direct supervised learning approaches less effective (Tauxe & Yamazaki, 2015). To overcome this, exploratory data analysis (EDA) was conducted to identify key trends and variations in the dataset, followed by an unsupervised learning approach to detect inherent patterns within geological features. These insights helped refine the dataset and guide the generation of synthetic lithology data to augment real observations.

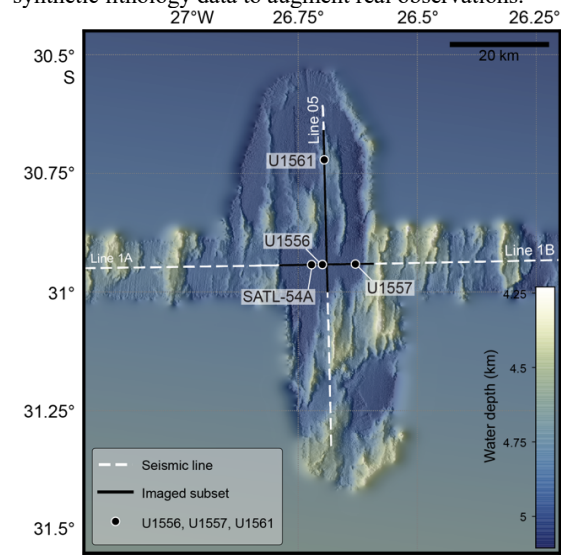

Figure 1: Sites U1556, U1557, and U1561 bathymetry (Christeson and Reece, 2020). Seismic reflection profiles were acquired during CREST cruise (Reece et al., 2016).

The creation of synthetic data and/or training examples (input-target pairs) plays a crucial role in addressing the challenges posed by unbalanced and incomplete datasets. By generating lithological data that aligned with realistic geophysical and geochemical properties, a more robust training dataset for machine learning models was proposed (de Leeuw et al., 2019). Traditional statistical methods have been used along with ML (Karmakar et al., 2018) for IODP

## Machine Learning for Lithology Interpretation

data analysis and their effectiveness should be the basis for development of more sophisticated AI/ML techniques. The objectives of the present study; (i) to develop supervised ML for IODP data/well log data (ii) to develop unsupervised ML to justify the number of clusters that support the data-cluster naturally. (iii) To compare results of Supervised and unsupervised ML.

Here, various ML models, including Logistic Regression, Decision Trees, Gradient Boosting, Support Vector Machines (SVM), k-Nearest Neighbors (KNN), and Multi-Layer Perceptron (MLP), were built and calibrated Upon successful training and calibration, the ML is applied to real IODP 390/393 dataset.

**Methodology**

A combination of exploratory data analysis (EDA), unsupervised clustering, synthetic data generation, and supervised machine learning was leveraged to enhance lithological classification from IODP well log data. The workflow consisted of key steps: data preprocessing and exploratory analysis, unsupervised clustering for pattern recognition, synthetic data generation, training and evaluation of machine-learning models, and final prediction of real IODP data.

Well-log data from IODP Expeditions 390/393 included measurements such as bulk density (RHOM), compressional velocity (VELP), gamma-ray (HSGR), porosity (APLC), and resistivity (RT_HRLT). The dataset contained missing values and noise, which were addressed using interpolation, outlier removal, and normalization techniques. EDA revealed significant variability, with RHOM values ranging from 1.30 to 2.98 g/cc and VELP values between 2.50 and 7.19 km/s. Unsupervised clustering using k-means and Gaussian Mixture Models (GMM) helped identify natural groupings in the dataset, which were validated against geological knowledge.

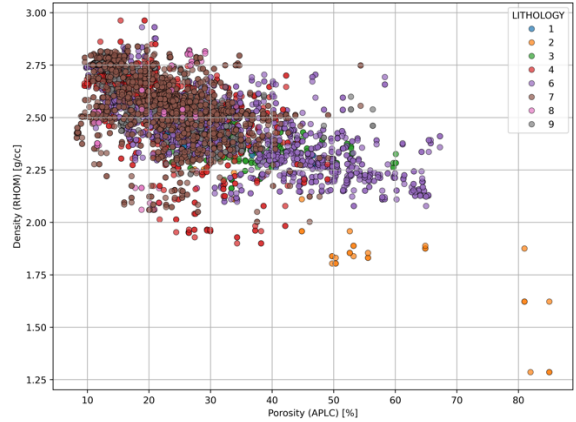

Figure 3: Crossplot of Apparent Porosity (APLC) vs. Bulk Density (RHOM) from IODP expedition data.

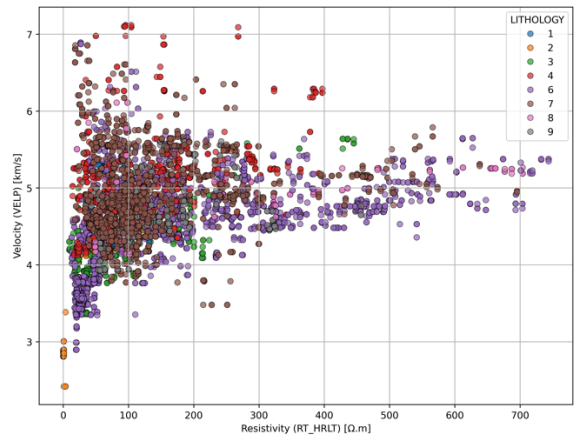

Figure 4: Crossplot of Thermal Resistivity (RT_HRLT) vs. P-wave Velocity (VELP) from IODP expedition data.

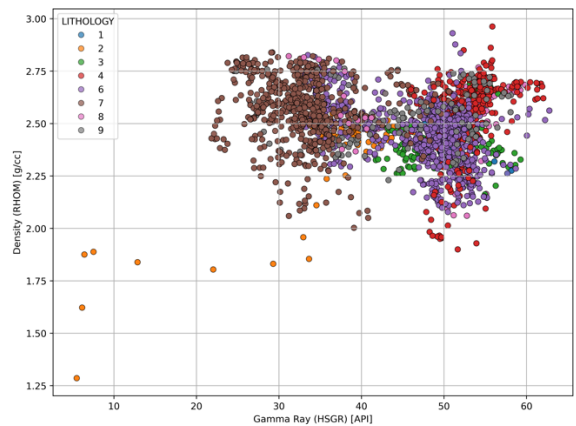

Figure 5: Crossplot of Natural Gamma Ray (HSGR) vs. Bulk Density (RHOM) from IODP expedition data.

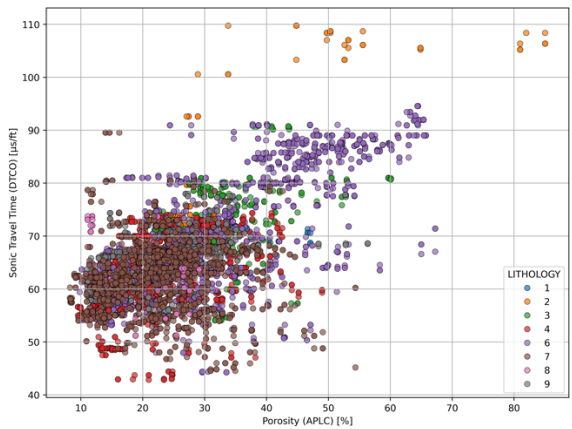

Figure 2: Crossplot of Apparent Porosity (APLC) vs Compressional Sonic Travel Time (DTCO) from IODP expedition data.

# Machine Learning for Lithology Interpretation

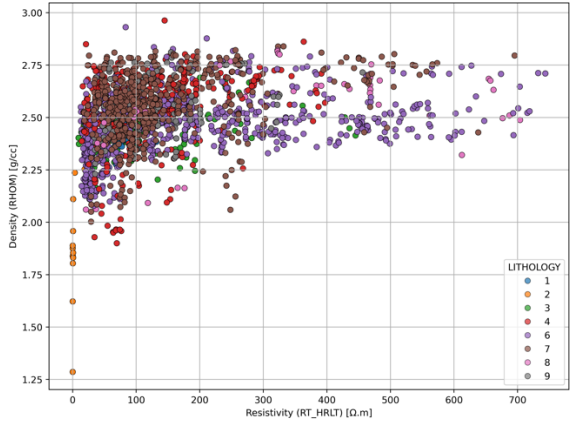

Figure 6: Crossplot of Thermal Resistivity (RT_HRLT) vs. Bulk Density (RHOM) from IODP expedition data.

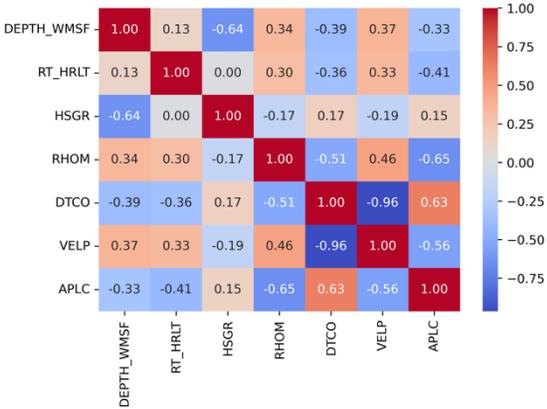

Figure 7: Correlation matrix of IODP expedition data

Figures 2–6 present crossplots of key geophysical properties derived from IODP expedition data, illustrating relationships between apparent porosity, sonic travel time, bulk density, thermal resistivity, P-wave velocity, and natural gamma ray measurements. These plots reveal significant overlap among multiple lithologies, indicating that distinct rock types exhibit similar physical property distributions. This overlap complicates direct lithological classification, as traditional threshold-based approaches struggle to delineate clear boundaries between lithologies. Consequently, a machine learning-driven approach becomes essential to capture subtle patterns and improve classification accuracy. Figure 7 shows the correlation matrix of well-log features,

Table 1: Statistical summary of the synthetic dataset, showing key descriptive statistics (mean, standard deviation, minimum, and maximum) for each feature

| Feature | Count | Mean | Std Dev | Min | Max |
|---|---|---|---|---|---|
| RT_HRLT | 4000 | 242.1 | 183.2 | 0.52 | 749.7 |
| DTCO | 4000 | 68.66 | 13.49 | 42.01 | 109.9 |
| VELP | 4000 | 4.80 | 0.98 | 2.51 | 7.19 |
| RHOM | 4000 | 2.38 | 0.30 | 1.30 | 2.98 |
| HSGR | 4000 | 46.52 | 12.11 | 6.11 | 69.90 |
| APLC | 4000 | 34.76 | 15.94 | 8.27 | 84.86 |

highlighting strong negative correlations between DTCO and VELP, as well as moderate relationships among other geophysical properties.

To address data imbalance and improve model generalization, synthetic well log data representing various lithologies were generated. Synthetic data distributions were sampled to align with real data trends, ensuring geologically plausible values. The key properties (RHOM, VELP, DTCO, HSGR, RT_HRLT, and APLC) were sampled within realistic ranges, introducing randomness to mimic natural variations.

Table 2 presents the synthetic lithology dataset, which defines the property ranges for key geophysical parameters: bulk density (RHOM), P-wave velocity (VELP), compressional wave transit time (DTCO), gamma-ray response (HSGR), high-resolution resistivity (RT_HRLT), and apparent porosity (APLC). These lithological categories ranging from breccia with cement to pillow lavas with feeder dikes represent characteristic rock types encountered in IODP Expeditions 390/393. The dataset was designed to simulate realistic variations observed in well log data, aiding in model training and evaluation for lithological classification.

Multiple supervised learning models were trained, including Decision Tree, Gradient Boosting,. Model performance was assessed using accuracy, F1-score, and confusion matrices (Pedregosa et al., 2011). The Decision Tree and Gradient Boosting models demonstrated superior classification

Table 2: Synthetic lithology dataset with property ranges for RHOM, VELP, DTCO, HSGR, RT_HRLT, and APLC.

| Property | RHOM (g/cc) | VELP (km/s) | DTCO (µs/ft) | HSGR (API) | RT_HRLT (Ω.m) | APLC (%) |
|---|---|---|---|---|---|---|
| Breccia with cement | 2.20 - 2.65 | 4.10 - 5.55 | 56.00 - 72.50 | 45.00 - 62.00 | 50 - 140 | 19 - 50 |
| Sedimentary breccia with matrix | 1.30 - 2.55 | 2.50 - 4.60 | 65.00 - 110.00 | 6.00 - 55.00 | 0.10 - 130 | 20 - 85 |
| Hyaloclastite breccia with pillows | 2.15 - 2.70 | 3.30 - 5.70 | 55.00 - 91.00 | 42.00 - 60.00 | 8 - 450 | 15 - 60 |
| Thick pillow flows and/or massive flows | 1.88 - 2.98 | 4.00 - 7.20 | 42.00 - 75.00 | 45.00 - 70.00 | 12 - 400 | 9 - 50 |
| Pillow lavas with breccia | 2.05 - 2.95 | 2.85 - 6.95 | 43.00 - 95.00 | 28.00 - 63.00 | 12 - 750 | 9 - 70 |
| Pillow lavas | 2.00 - 2.85 | 3.40 - 6.90 | 44.00 - 90.00 | 24.00 - 55.00 | 18. - 700 | 8 - 55 |
| Massive flow(s) | 2.00 - 2.85 | 4.00 - 6.20 | 50.00 - 75.00 | 35.00 - 62.00 | 20 - 710 | 10 - 30 |
| Pillow lavas with feeder dikes | 2.22 - 2.82 | 3.90 - 5.60 | 54.00 - 78.00 | 25.00 - 65.00 | 34 - 460 | 10 - 58 |

# Machine Learning for Lithology Interpretation

Table 3: Performance metrics of different classification models on the synthetic dataset, evaluated using accuracy, precision, recall, and F1-score.

| Model | Accuracy | Precision | Recall | F1 Score |
|---|---|---|---|---|
| Logistic Regression | 0.8425 | 0.8433 | 0.8425 | 0.8426 |
| Decision Tree | 0.9950 | 0.9951 | 0.9950 | 0.9951 |
| Gradient Boosting | 0.9950 | 0.9951 | 0.9950 | 0.9951 |
| SVM | 0.9625 | 0.9629 | 0.9625 | 0.9625 |
| KNN | 0.9100 | 0.9156 | 0.9100 | 0.9083 |
| MLP Classifier | 0.9900 | 0.9902 | 0.9900 | 0.9900 |

Table 4: Performance comparison of various unsupervised clustering models using Adjusted Rand Index (ARI) and Normalized Mutual Information (NMI) scores.

| Clustering Model | ARI | NMI |
|---|---|---|
| Spectral Clustering | 0.2165 | 0.3023 |
| Agglomerative Clustering | 0.1809 | 0.3220 |
| K-means Clustering | 0.1675 | 0.2943 |
| Gaussian Mixture Model (GMM) | 0.1436 | 0.2462 |
| DBSCAN | -0.0006 | 0.1054 |

capabilities with an accuracy of 0.9950. The feature importance analysis identified VELP and RHOM as the most influential parameters in lithology classification.

The best-performing model was applied to the original IODP dataset to classify lithology in previously unlabeled intervals. To assess its effectiveness, the predictions were compared with geological descriptions and core sample interpretations. The results demonstrated a strong correlation between the model's classifications and expected geological patterns, highlighting the potential of machine learning for lithological interpretation. Table 3 summarizes the performance metrics of various models, with the Decision Tree and Gradient Boosting classifiers achieving the highest accuracy (99.50%), followed closely by the Multi-Layer Perceptron (MLP) classifier (99.00%) and the Support Vector Machine (SVM) (96.25%). Additionally, Table 4 presents the performance of various unsupervised clustering models, assessing their ability to group lithological units based on inherent patterns in the dataset (Jha et al., 2021). Among the clustering methods, Spectral Clustering achieved the highest ARI (0.2165) and Agglomerative Clustering obtained the highest NMI (0.3220), suggesting their superior capability in capturing lithological structures compared to other clustering techniques. Spectral Clustering performed the best among unsupervised methods because it effectively captures nonlinear relationships and complex cluster structures in the dataset by leveraging graph-based similarity measures, making it well-suited for identifying lithological boundaries in high-dimensional well log data.

**Conclusion**

A data-driven approach integrating machine learning with IODP well log data improved lithological classification. Unsupervised clustering revealed hidden patterns, aiding synthetic data generation, which enhanced model training. The best model, trained on synthetic data, was selected for real IODP data analysis, demonstrating the potential of automated subsurface characterization.

Figure 8 illustrates the effectiveness of different clustering methods. The Decision Tree model captures major lithological transitions, while Spectral Clustering identifies finer-scale variations. This comparison highlights the trade-off between model interpretability and geological sensitivity. Future work will focus on refining synthetic data, incorporating geological constraints, and exploring hybrid modelling approaches.

Future work will focus on refining synthetic data by incorporating additional geological constraints to enhance realism and variability. Additionally, integrating domain knowledge into machine learning models through physics-informed approaches and hybrid modelling techniques could improve classification accuracy. Expanding the dataset with more labelled samples and exploring deep learning architectures may further enhance the robustness and generalizability of the models.

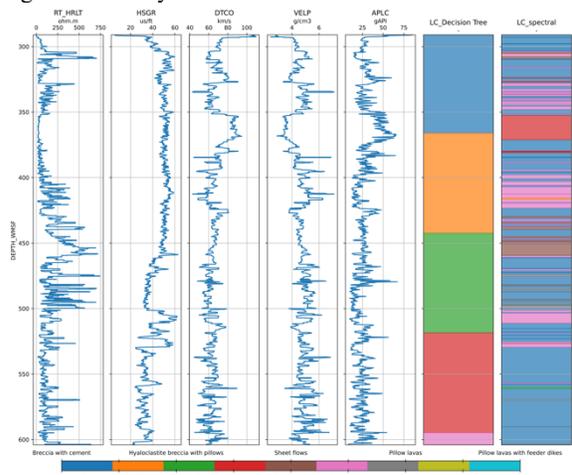

Figure 8: Well log analysis with lithology classification. The Decision Tree model shows well-defined lithological boundaries, while Spectral Clustering captures finer variations, highlighting transitions between lithologies.

**Acknowledgement**

We express our gratitude to the IODP for providing access to well log datasets from Expeditions 390/393. We also acknowledge the contributions of geoscientists, whose previous work on well-log interpretation has guided this study. Special thanks to IIT ISM Dhanbad for providing the computational resources and support. Finally, we appreciate the open-source community for developing machine-learning frameworks that facilitate this research.